\documentclass[10pt,twocolumn,letterpaper]{article}

\usepackage{iccv}
\usepackage{times}
\usepackage{epsfig}
\usepackage{graphicx}
\usepackage{amsmath}
\usepackage{amssymb}

\usepackage{times}
\usepackage{epsfig}
\usepackage{graphicx}
\usepackage{amsmath}
\usepackage{amssymb}
\usepackage{soul}
\usepackage{adjustbox}
\usepackage{threeparttable}
\usepackage{multirow}
\usepackage{xcolor}
\usepackage[breaklinks=true,bookmarks=false]{hyperref}

\iccvfinalcopy 

\def\iccvPaperID{****} 

\ificcvfinal\fi

\begin{document}

\title{VTGAN: Semi-supervised Retinal Image Synthesis and Disease Prediction using Vision Transformers}

\author{Sharif Amit Kamran\\
University of Nevada, Reno\\
{\tt\small skamran@nevada.unr.edu}
\and
Khondker Fariha Hossain\\
University of Nevada, Reno\\
{\tt\small  khondkerfarihah@nevada.unr.edu}
\and
Alireza Tavakkoli\\
University of Nevada, Reno\\
{\tt\small  tavakkol@unr.edu}
\and
Stewart Lee Zuckerbrod\\
Houston Eye Associates\\
{\tt\small  szuckerbrod@houstoneye.com}
\and
Salah A. Baker\\
University of Nevada, Reno\\
{\tt\small  sabubaker@med.unr.edu}
}

\maketitle
\ificcvfinal\thispagestyle{empty}\fi

\begin{abstract}
In Fluorescein Angiography (FA), an exogenous dye is injected in the bloodstream to image the vascular structure of the retina. The injected dye can cause adverse reactions such as nausea, vomiting, anaphylactic shock, and even death. In contrast, color fundus imaging is a non-invasive technique used for photographing the retina but does not have sufficient fidelity for capturing its vascular structure. The only non-invasive method for capturing retinal vasculature is optical coherence tomography-angiography (OCTA). However, OCTA equipment is quite expensive, and stable imaging is limited to small areas on the retina. In this paper, we propose a novel conditional generative adversarial network (GAN) capable of simultaneously synthesizing FA images from fundus photographs while predicting retinal degeneration. The proposed system has the benefit of addressing the problem of imaging retinal vasculature in a non-invasive manner as well as predicting the existence of retinal abnormalities. We use a semi-supervised approach to train our GAN using multiple weighted losses on different modalities of data. Our experiments validate that the proposed architecture exceeds recent state-of-the-art generative networks for fundus-to-angiography synthesis. Moreover, our vision transformer-based discriminators generalize quite well on out-of-distribution data sets for retinal disease prediction. 
\end{abstract}
\section{Introduction}
Fluorescein Angiography (FA) combined with retinal funduscopy is a standard tool for diagnosing various retinal vascular abnormalities and degenerative conditions \cite{mary2016retinal}. In Fluorescein Angiography, a fluorescent fluid is injected into the blood stream and becomes visible 8 to 10 minutes after insertion. The dosage depends on the viscosity of the dye, age of the patient, and the retina vascular structure \cite{mandava2004fluorescein}. In general, the procedure is safe, but there have been reported cases of allergic reactions, ranging from nausea and vomiting to anaphylactic shock and death \cite{lira2007adverse,kwan2006fluorescein,kwiterovich1991frequency}. The risk associated with FA signifies the need for non-invasive mechanisms to assess retinal vascular structure.

Various automated systems incorporating image processing and machine learning techniques have been proposed for diagnosing retinal abnormalities and degenerative diseases from fundus photographs \cite{gurudath2014machine,fu2018disc,poplin2018prediction,lira2007adverse}. Currently, there are no inexpensive imaging alternative for imaging the vascular structure of the retina. Although OCTA \cite{huang1991optical} is capable of mapping the retinal subspace in 3D, the equipment is expensive and its highest resolutions are only achievable on very small areas. Consequently, it is imperative to develop non-invasive and inexpensive techniques for measuring retinal vasculature to circumvent the risks associated with the existing invasive procedures. The first contribution of this paper is a novel deep learning architecture for producing retinal vascular images (i.e., FA images) from non-invasive fundus photographs. 

A major obstacle in using machine learning architectures for ophthalmic applications is the lack of publicly available data -- in our work Fluorescein Angiography data \cite{kamran2019optic,kamran2020improving}.  Only a handful of machine learning systems have been proposed to predict disease using FA images. However, these systems are trained and tested on privately-held data set. Pan \etal. \cite{pan2020multi} utilized three such pre-trained architectures on 4067 privately curated images for Abnormal and Normal FA prediction.  Similarly, Li \etal. \cite{li2020automated} incorporated the encoder of U-Net \cite{ronneberger2015u} for classifying different levels of degradation for 3935 privately held Ultra-widefield FA images. To our knowledge, the only two available public data sets have only 59, and 70 FA images in total  \cite{hajeb2012diabetic,alipour2014new}. The second contribution of this paper is a semi-supervised approach in conjunction with a vision transformer architecture to address the challenges in training models that learn pathologies from such small amounts of data. 

\begin{figure*}[ht]
    \centering
    \includegraphics[width=0.9\linewidth]{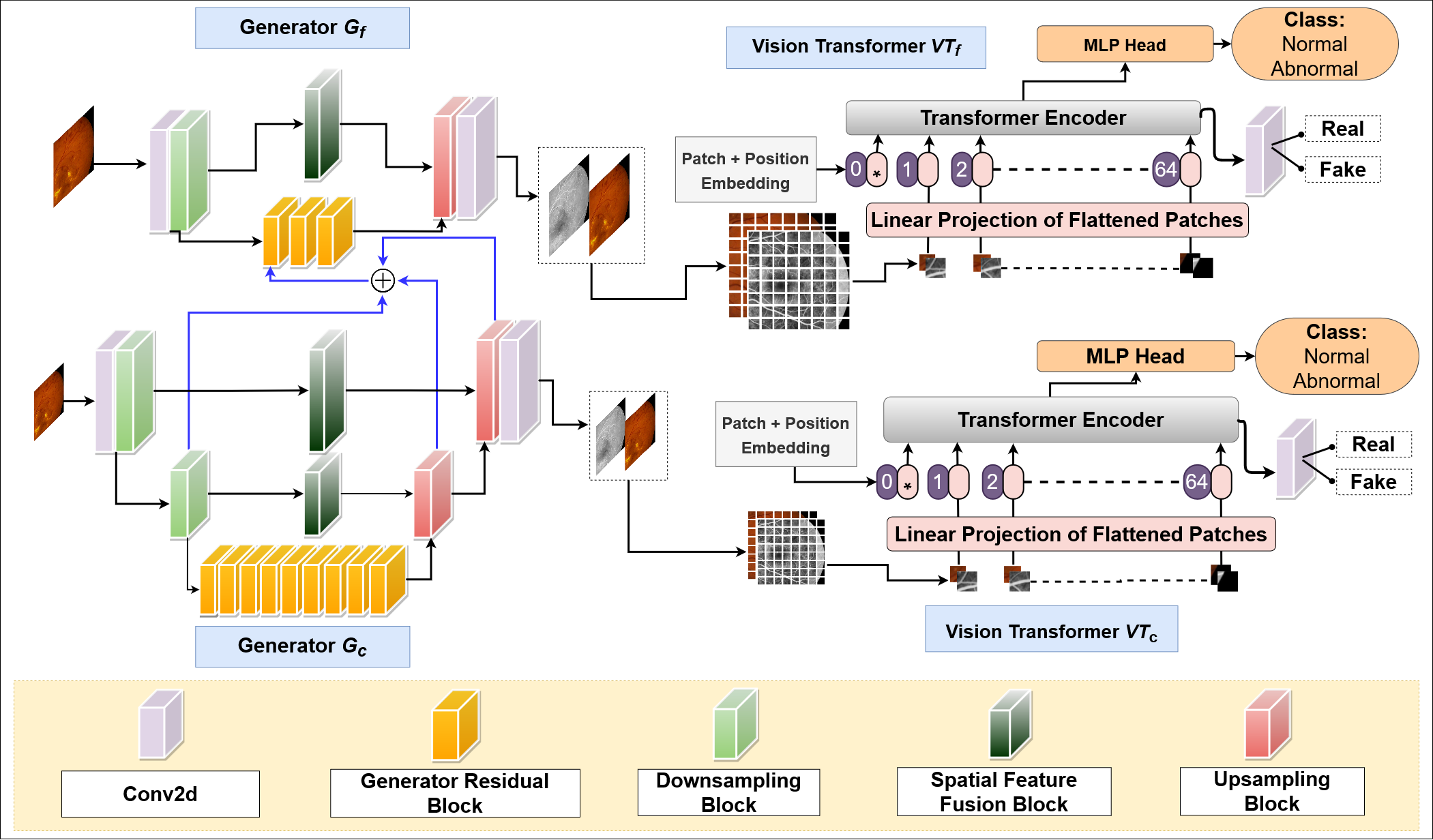}
    \caption{VTGAN consists of Coarse and Fine generators $G_f,G_c$ and Vision Transformers as discriminators $VT_f,VT_c$. The generators take fundus as input and synthesize angiograms. Whereas the vision transformers take patches of the concatenated fundus and angiograms as input and outputs, i) feature-map for adversarial example detection, and ii) image-level classification for Fluorescein angiograms. The generators consist of Downsampling, Upsampling, Spatial Feature Fusion, Residual blocks and multi-scale feature summation (\textcolor{blue}{Blue arrow}). The Vision transformers consist of Transformer Encoder, Multi-layer perceptron (MLP) head, and adversarial feature output block.}
    \label{fig1}
\end{figure*}

In particular,  we introduce VTGAN, a semi-supervised conditional GAN that can simultaneously produce the retinal vascular structure (i.e. FA images) from fundus photographs, while differentiating between healthy and abnormal retina. The proposed architecture incorporates a novel discriminator based on vision transformers for both patch-level adversarial image detection and image-level disease classification. Specifically, we utilize a novel embedding feature loss incorporating patch and positional information to find similarities between angiography and fundus images for better disease prediction. For qualitative assessment, we compare our proposed architecture with recent state-of-the-art fundus-to-angiogram synthesis architectures ~\cite{kamran2020fundus2angio, kamran2020attention2angiogan, wang2018high,kim2019u,choi2020stargan}. For quantitative evaluation, we use Frechet inception Distance (FID)~\cite{heusel2017gans} and Kernel Inception Distance (KID)~\cite{binkowski2018demystifying} for quantifying image features and measuring structural similarity. To validate the discriminator's robustness, we use standard metrics for out-of-distribution classification. 

\section{Related Work}
Generative adversarial networks (GAN) have become a staple for image-to-image synthesis \cite{chen2018sketchygan,sangkloy2017scribbler}, inpainting \cite{zhu2016generative,dekel2018sparse} and style transfer \cite{wang2018high,xian2018texturegan}. By a combination of multi-scale architectures, these networks can detect and learn fine and coarse features from images \cite{burt1983laplacian,brown2003recognising}.
This approach, in turn, can be employed for both conditional \cite{huang2017stacked,denton2015deep} and unconditional variants of GAN architectures \cite{chen2017photographic,zhang2017stackgan}. Generative networks have also seen success in computer tomography (CT), magnetic resonance imaging (MRI), and X-ray for image segmentation, augmentation and cross-domain information fusion tasks \cite{nie2017medical,nie2018medical,shin2018medical,waheed2020covidgan}. 
Recently, GAN models are utilized for synthesizing FA images from fundus photographs  \cite{tavakkoli2020novel, kamran2020attention2angiogan}. These models employ two generators for coarse and fine image generation. The generators were trained on randomly cropped patches of different scales, while multi-scale discriminators are used to discern local and global features from images. Usually, a discriminator incorporates a linear output for classifying among real or fake (generated) images. In addition, these models utilize the PatchGAN architecture \cite{li2016precomputed} as a discriminator for adversarial and real image classification, by producing the output as a feature map of size $N\times N$. Incorporating these ideas, the state-of-the-art models employ an architecture that can retain global information like the shape of optic-disc, contrast, and local features like venular structures, arteries, and microaneurysm \cite{tavakkoli2020novel, kamran2020attention2angiogan}. However, the problem with this approach is that the discriminator works on patch-level information. As a result, the cohesive relationship between global and local features is lost while generating FA from crops of fundus images. 

The advent of Vision Transformers (ViT) have improved the performance of state-of-the-art architectures in image classification tasks.  The premise of ViT is that image pixels contain inherent spatial coherencies. Therefore, utilizing an image as a sequence of non-overlapping patches and incorporating them into a transformer can better extract intrinsic and spatial features \cite{vaswani2017attention}. Unlike image generative pre-trained models such as iGPT \cite{chen2020generative}, which apply transformers to obtain image-level features, ViT works on patch-level features and incorporates their positions by utilizing embedding layers \cite{khan2021transformers,chen2020generative}. 

Vision Transformers retain the cohesiveness of coarse and fine features by utilizing the position of each patch. By mapping this information for $N\times N$ patches into an $N\times N$ feature-map, we propose a novel architecture based on vision transformers, called VTGAN and illustrated in Fig.~\ref{fig1}. Moreover, we extend VTGAN's capability by adding a multi layer perceptron (MLP) head for classifying Abnormal and Normal FA images in a semi-supervised manner. This architecture is a significant contribution that addresses training models for FA image classification task based on limited amounts of publicly available data. Qualitative and quantitative evaluations demonstrate that VTGAN surpasses other state-of-the-architectures both in terms of synthesis and classification tasks. Additionally, we demonstrate that the produce FA images by VTGAN are such high quality that expert ophthalmologists cannot reliably identify fake image from real ones.

\section{Proposed Methodology}
This paper proposes a vision-transformer-based generative adversarial network (GAN) consisting of residual, spatial feature fusion, upsampling and downsampling blocks for generators, and transformer encoder blocks for discriminators. We incorporate multiple losses for generating vivid fluorescein angiography images from normal and abnormal fundus photographs for training. In the first section, we elaborate coarse and fine generators in section \ref{subsec:generators}. We detail each distinct blocks in sections \ref{subsec:encdec}, \ref{subsec:residualblock}, and \ref{subsec:attention}. We then elaborate on our newly proposed vision-transformer-based discriminators and the interrelationship between the generators and discriminators to establish the overall generative network in section \ref{subsec:discriminators}. Finally, in section \ref{subsec:objective} we discuss the associated loss functions and their weight multipliers for each distinct architecture that forms the proposed model. 
\begin{figure}[htb]
    \centering
    \includegraphics[width=0.9\linewidth]{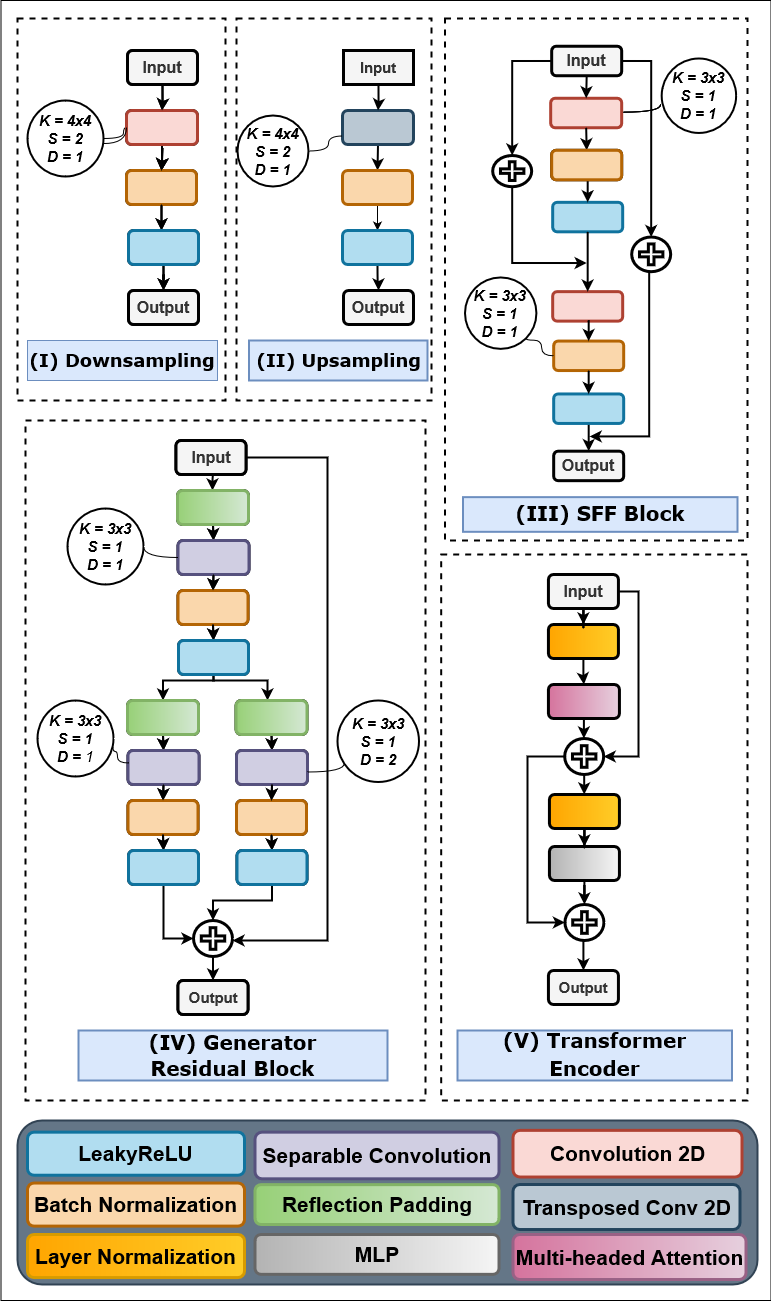}
    \caption{Distinct building blocks of our proposed Generative network comprising of (i) Downsampling, (ii) Upsampling, (c) Spatial Feature Fusion (SFF) block, (d) Generator Residual Block and (e) Transformer Encoder. Here,  K = for kernel size, S = stride  and D = Dilation rate. }\label{fig2}
\end{figure}

\subsection{Multi-scale Generators}
\label{subsec:generators}
In order to capture large and fine-scale features to produces realistic vascular images, we combine multi-scale coarse and fine generators.  We adopt two generators ($G_{f}$ and $G_{c}$), as visualized in Fig.~\ref{fig1} in our architecture. $G_{f}$ synthesizes local features such as arteries and venules. Conversely, $G_{c}$ translates global features such as large blood vessels, optic disc, and overall contrast and illumination. The generators consist of multiple downsampling, upsampling, spatial feature fusion, residual blocks, and a multi-scale feature summation block (\textcolor{blue}{Blue arrow}) between the $G_{f}$ and $G_{c}$. The input and output dimension of $G_{f}$ are $512\times 512$, while $G_{c}$ has half the dimension of $G_{f}$ ($256\times 256$). Furthermore, the $G_{c}$ generates $256\times 256\times 64$ size feature vector, which is element-wise added with one of the intermediate layers of the $G_{f}$ generator. Both $G_{c}$ and $G_{f}$ take fundus images and synthesize FA images. The detailed arrangement is visualized in Fig.~\ref{fig1}.

\subsection{Downsampling and Upsampling Blocks}
\label{subsec:encdec}
We use, as generators, auto-encoders comprising of multiple downsampling and upsampling blocks for feature extraction. A single downsampling block contains a convolution layer, a batch-norm layer \cite{ioffe2015batch} and a Leaky-ReLU activation function successively and is given in Fig.~\ref{fig2}(i). In contrast, an upsampling block consists of a transposed convolution layer, batch-norm \cite{ioffe2015batch}, and Leaky-ReLU activation layer consecutively and is illustrated in  Fig.~\ref{fig2}(ii).  We use the downsampling block twice in $G_{c}$, followed by nine successive residual identity blocks. Finally, the upsampling blocks are used $2\times$ again to make the spatial output the same as the input. For $G_{f}$, we utilize the downsampling once, and after three consecutive residual blocks, a single upsampling block is employed to get the same spatial output as the input. We use kernel size, $k=3$, and stride, $s=2$ for convolution 2D and transposed convolution 2D layers. 

\subsection{Generator's Residual Blocks}
\label{subsec:residualblock}
For spatial and depth feature extraction, residual blocks have become the fundamental building blocks for image-to-image translation, inpainting, and style transfer tasks \cite{shaham2019singan,wang2018high,choi2020stargan,choi2018stargan,park2019semantic}.  The original residual block consisted of successive convolutions and a skip connection linking the pre-residual layer with the post-residual layer \cite{he2016deep}. Regular convolution operations are computationally inefficient and fail to retain accurate spatial and depth information, unlike separable convolution \cite{chollet2017xception}. Separable convolution incorporates a depth-wise convolution layer followed by a point-wise convolution. Consequently, it obtains and retains depth and spatial features better. We use residual blocks for our generators, as illustrated in Fig.~\ref{fig2}(iv). The residual block consists of Reflection padding, Separable Convolution, Batch-Norm, and Leaky-ReLU layers followed by two branches of the same repetitive layers. The main difference is, one branch consists of a dilation rate of $d=1$ and the other with a dilation rate, $d=2$ in the separable convolution layers. We use, kernel size, $k=3$ and stride, $s=1$. The output from the two branches and the skip-connection successively element-wise summed in the final output.

\subsection{Spatial Feature Fusion Block}
\label{subsec:attention}
In this section, we elaborate on the spatial feature fusion (SFF) block, as illustrated in Fig.~\ref{fig2}(iii). The block consists of two residual units with Convolution, Batch-Norm, Leaky-ReLU layer successively. The convolution layer have kernel, $k=3$, and stride, $s=1$. Additionally, there are two skip connections, one going from the input and element-wise summed to the first residual unit's output. The next one comes out of the input layer and is added with the last residual unit's output. We use spatial feature fusion block for combining spatial features from the bottom layers with the topmost layers of the architecture, as visualized in Fig.~\ref{fig1}. $G_{c}$ comprises two SFF blocks that connect each of the two downsampling blocks with the two upsampling blocks successively. In contrast, $G_{f}$ has only one SFF block between the single downsampling and upsampling block. The reason behind incorporating the SFF block is to extract and retain spatial information that is otherwise lost due to consecutive downsampling and upsampling. As a result, we can combine these features with the learned features of the later layers of the network to get an accurate approximation, as seen in similar GAN architectures \cite{zhang2019self,chen2018attention}. 

\subsection{Vision-transformer as Discriminators}
\label{subsec:discriminators}
GAN discriminators require adapting to local and global information changes for differentiating real and fake images. To alleviate this inherent problem, we need a heavy architecture with a large number of parameters.  In contrast, convolution with a large receptive field can be employed for obtaining multi-scale features but can cause overfitting on training data. To resolve this problem, we propose a new Vision Transformer-based Markovian discriminator, in the same vein as PatchGAN \cite{li2016precomputed}. We use eight vision transformer encoders consisting of a multi-headed attention layer and multi-layer perceptron (MLP) block. The Layer Normalization layer precedes each block, and residual skip connection is added to the output from the input. The block is visualized in Fig.~\ref{fig2}(v). To handle 2D images of $512\times512$, we reshape the images into a sequence of flattened 2D patches with resolution $64\times64$. By doing so, we end up having 64 patches in total. The Transformer uses a constant latent vector size of $D=64$ through all of its layers, so we flatten the patches and map to  $D$ dimensions with a trainable linear projection. The output of this projection is called the patch embeddings as mentioned in \cite{dosovitskiy2020image}.  Position embeddings are added to the patch embeddings to preserve positional information. We use regular learnable 1D position embeddings in a similar manner to Dosovitskiy \etal \cite{dosovitskiy2020image}. For multi-headed attention, we use $n=4$ heads. For MLP blocks, we use two dense layers with features size $h=[128,64]$, each succeeded by a GeLU activation \cite{hendrycks2016gaussian} and a dropout of 0.1. Contrarily, our vision transformer has two outputs, an MLP head, and a Convolutional layer, as illustrated in Fig.~\ref{fig1}. The MLP head has two output hidden units for FA image classification (Abnormal and Normal). In contrast, the convolution layer outputs a feature map of $64\times64$ for classifying each patch in the original image as Real or Fake.

We use two vision transformer-based discriminators that incorporate identical structures but operate at two different scales. We term the two discriminators as, $VT_{f}$ and $VT_{c}$ as visualized in Fig.~\ref{fig1}.  The coarse angiograms and fundus are resized to $256\times256$ by a factor of $2$ using the Lanczos filter~\cite{duchon1979lanczos}.  Both discriminators have identical transformer encoder and output layers (in Fig.~\ref{fig2}(v)). 

$VT_{c}$ dictates $G_{c}$ to extract more global features such as the fovea, brightness, optic disc, and color sensitivity. Contrarily,  the $VT_{f}$ steers the fine generator, $G_{f}$ to synthesize more accurate fine features such as small vasculature, venules, exudates,  arteries. Consequently, we fuse learnable elements from both generators while training them with their paired vision transformer-based discriminators.

\subsection{Adversarial Cost Functions}
\label{subsec:objective}

Our whole network's objective function can be contrived as Eq.~\ref{eq1}. So it's a problem spanning multiple distinct objectives for the generator and discriminator networks, where our target is to maximize $VT_{f}$ and $VT_{c}$'s loss while minimizing the $G_{f}$ and $G_{c}$'s loss. 
\begin{equation}
    \min \limits_{G_{f},G_{c}} \max \limits_{VT_{f},VT_{c}}  \mathcal{L}_{adv}(G_{f},G_{c}, VT_{f},VT_{c})
    \label{eq1}
\end{equation} 

We employ Hinge-Loss \cite{zhang2019self,lim2017geometric} to train our model in an adversarial setting, as illustrated in Eq.~\ref{eq2} and Eq.~\ref{eq3}. All the fundus images and their corresponding angiogram pairs are normalized to $[-1,1]$, and we employ $tanh$ as the output activation for the adversarial feature-map.  We multiple $\lambda_{adv}(G)$ with a weight multiplier $\mathcal{L}_{adv}(G)$ and add $\lambda_{adv}(D)$ to the output as given in Eq.~\ref{eq4}. For classification we use $softmax$ activation after the MLP head output.  In Eq.~\ref{eq5} the categorical cross-entropy loss is given where $y$ is the real class and $\hat{y}$ is the predicted class.
\begin{multline}
    \mathcal{L}_{adv}(VT) = - \mathbb{E}_{x,y} \big[\ \min(0,-1+VT(x,y))\big]\ \\-  \mathbb{E}_{x} \big[\ \min(0,-1-VT(x,G(x))) \big]\
    \label{eq2}
\end{multline}
\begin{equation}
    \mathcal{L}_{adv}(G) = - \mathbb{E}_{x,y} \big[(VT(G(x),y))\big]\
    \label{eq3}
\end{equation}
\begin{equation}
    \mathcal{L}_{adv}(G,VT) = \mathcal{L}_{adv}(VT) + \lambda_{adv} (\mathcal{L}_{adv}(G)) 
    \label{eq4}
\end{equation}
\begin{equation}
    \mathcal{L}_{cce}(VT) = \mathbb{E}_{y,\hat{y}} \big[ - \sum_{i=1}^{k} y_i \log \hat{y_i}\big]\
    \label{eq5}
\end{equation}
Eq.~\ref{eq2} and Eq.~\ref{eq3} demonstrate $VT_{f}$ and $VT_{c}$ being trained on the real fundus ($x$) and FA ($y$) images  and then real fundus ($x$) and fake FA ($G(x)$) images in succession. We batch-wise train the vision transformers $VT_{f}$ and $VT_{c}$ for a couple of repetitions on images sampled randomly. Following that, $G_{c}$  and $G_{f}$ are trained on a batch of random images while freezing the weights the vision transformers, $VT_{c}$ and $VT_{f}$.

We utilize MSE loss and VGG-19 based perceptual loss ~\cite{johnson2016perceptual} for $G_{c}$ and $G_{f}$ as given in Eq.~\ref{eq6} and Eq.~\ref{eq7}. We ensure the synthesized images retain more realistic color, contrast, and vascular structure by employing these losses. For our discrimnators, $VT_{c}$ and $VT_{f}$ we incorporate a novel embedding feature loss as provided in Eq.~\ref{eq8}.

\begin{equation}
    \mathcal{L}_{MSE}(G) = \mathbb{E}_{x,y} \Vert G(x) - y \Vert^2
    \label{eq6}
\end{equation}
\begin{equation}
    \mathcal{L}_{perc}(G) = \mathbb{E}_{x,y} \sum_{i=1}^{k}\frac{1}{P} \Vert F_{VGG19}^{i}(y) - F_{VGG19}^{i}(G(x))\Vert
    \label{eq7}
\end{equation}
\begin{equation}
    \mathcal{L}_{ef}(G,VT) = \mathbb{E}_{x,y} \sum_{i=1}^{k}\frac{1}{Q} \Vert VT_{em}^{i}(x,y) - VT_{em}^{i}(x,G(x))\Vert
    \label{eq8}
\end{equation}

For synthesizing FA at different scales, we utilize the mean-squared-error loss $\mathcal{L}_{MSE}$ as shown in Eq.~\ref{eq6}. The difference between the real FA, $y$,  and the synthesized FA, $G(x)$ is calculated using this loss.  In contrast, the perceptual loss, $\mathcal{L}_{perc}$ is employed for findings similarities between real and fake FA as given in Eq.~\ref{eq7}.  Both of these images are pushed successively in VGG19~\cite{simonyan2014very} architecture, and then the resultant feature is subtracted.  Finally, the novel embedding feature loss is calculated by obtaining positional and patch features from the transformer encoder layers of $VT$ by inserting the real and synthesized FA succesively, as shown in Eq.~\ref{eq8}. Here, $P$ and $Q$ stand for the number of features extracted from VGG19 layers and the embedding layers of the transformer-encoder consecutively.

We combine Eq.~\ref{eq4}, \ref{eq5}, \ref{eq6}, \ref{eq7} and \ref{eq8} to configure our ultimate cost function as provided in Eq.~\ref{eq9}.
\begin{multline}
\min \limits_{G_{f},G_{c}} \big( \max \limits_{VT_{f},VT_{c}}  (\mathcal{L}_{adv}(G_{f},G_{c}, VT_{f},VT_{c})) + \\ \lambda_{MSE}\big[\ \mathcal{L}_{MSE}(G_{f},G_{c})\big]\ + 
\lambda_{ef}\big[\ \mathcal{L}_{ef}(G_{f},G_{c}, VT_{f},VT_{c})\big]\ \\ + \lambda_{perc}\big[\ \mathcal{L}_{perc}(G_{f},G_{c})\big]\  + \lambda_{cce}\big[\ \mathcal{L}_{cce}(VT_{f},VT_{c})\big]\ \big)
\label{eq9}
\end{multline}

$\lambda_{adv}$, $\lambda_{ef}$, $\lambda_{perc}$, $\lambda_{MSE}$, and $\lambda_{ccse}$ implies weights of the lossses, and priotizes a specific loss during training the networks. In our generative network, we emphasize more on $\mathcal{L}_{adv}(G)$, $\mathcal{L}_{MSE}$, $\mathcal{L}_{perc}$, $\mathcal{L}_{cce}$ , and thus we pick larger $\lambda$ values for these. 

\section{Experiments}
The following section outlines our proposed architectures experimentations and evaluations on quantitative \& qualitative metrics. First, we discuss the pre-processing pipeline for our dataset in Sec.~\ref{subsec:dataset}. Next, we detail our hyper-parameter selection and tuning in Sec.~\ref{subsec:hyper}. Moreover, we compare our model with other fundus-to-angio synthesis models with various qualitative metrics in Sec.~\ref{subsec:qual}. Lastly, in Sec.~\ref{subsec:quant} and Sec.~\ref{subsec:classification},  we examine the expert quantifications and the performance on the out-of-distribution dataset. 

\begin{figure}[t]
    \centering
    \includegraphics[width=\linewidth,height=13.5cm]{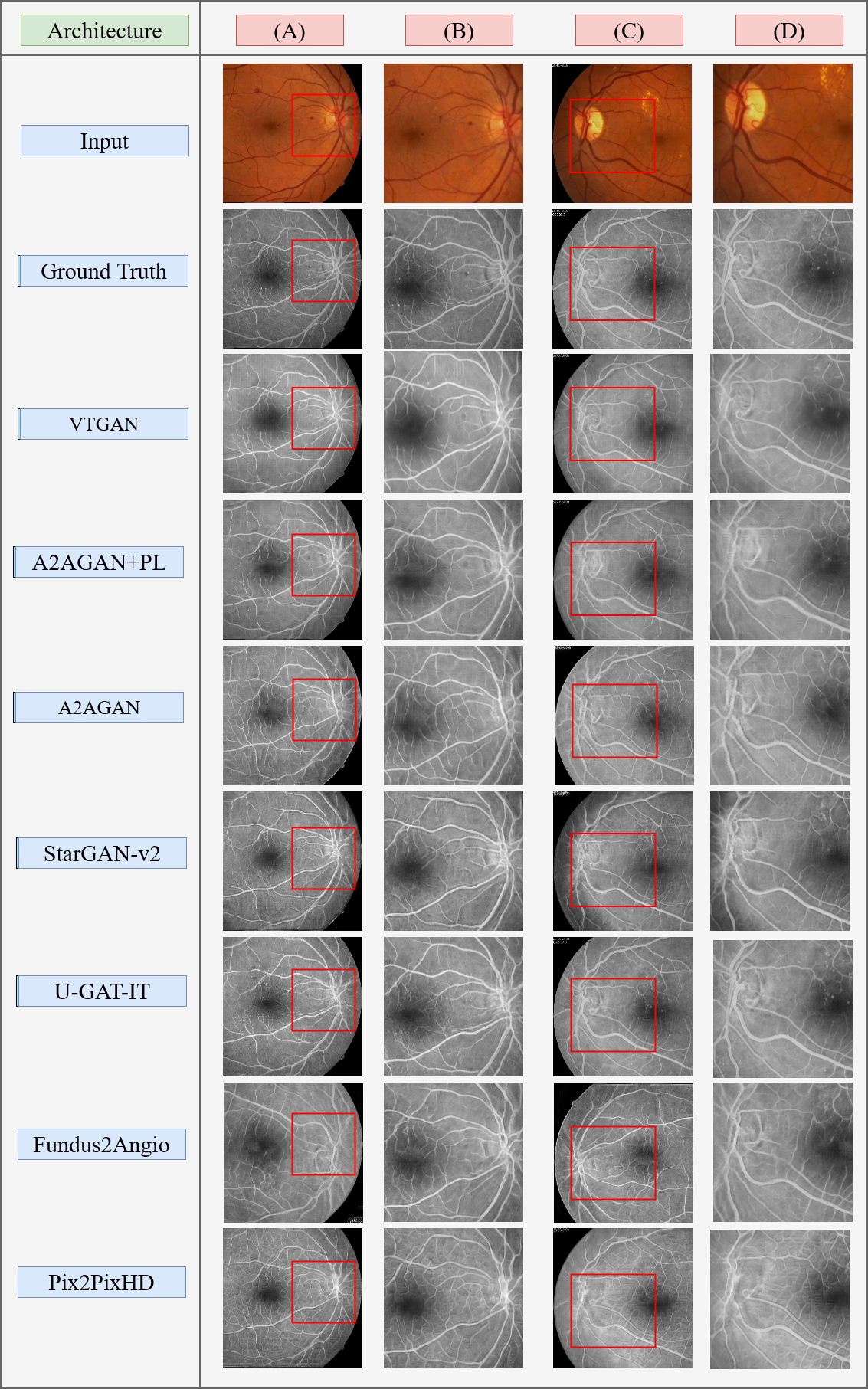}
    \caption{Relative differences in Fluroscein Angiograms (FA) synthesized by various generative networks. Columns (A) and (C) depict real fundus, real FA, and synthesized FA images. In contrast, column (B) and (D) displays zoomed-in red rectangle boxes from Column (A) \& (C), to show similar local vessels and venular arrangement.}
    \label{fig3}
\end{figure}
\begin{figure}[t]
    \centering
    \includegraphics[width=\linewidth,height=13.5cm]{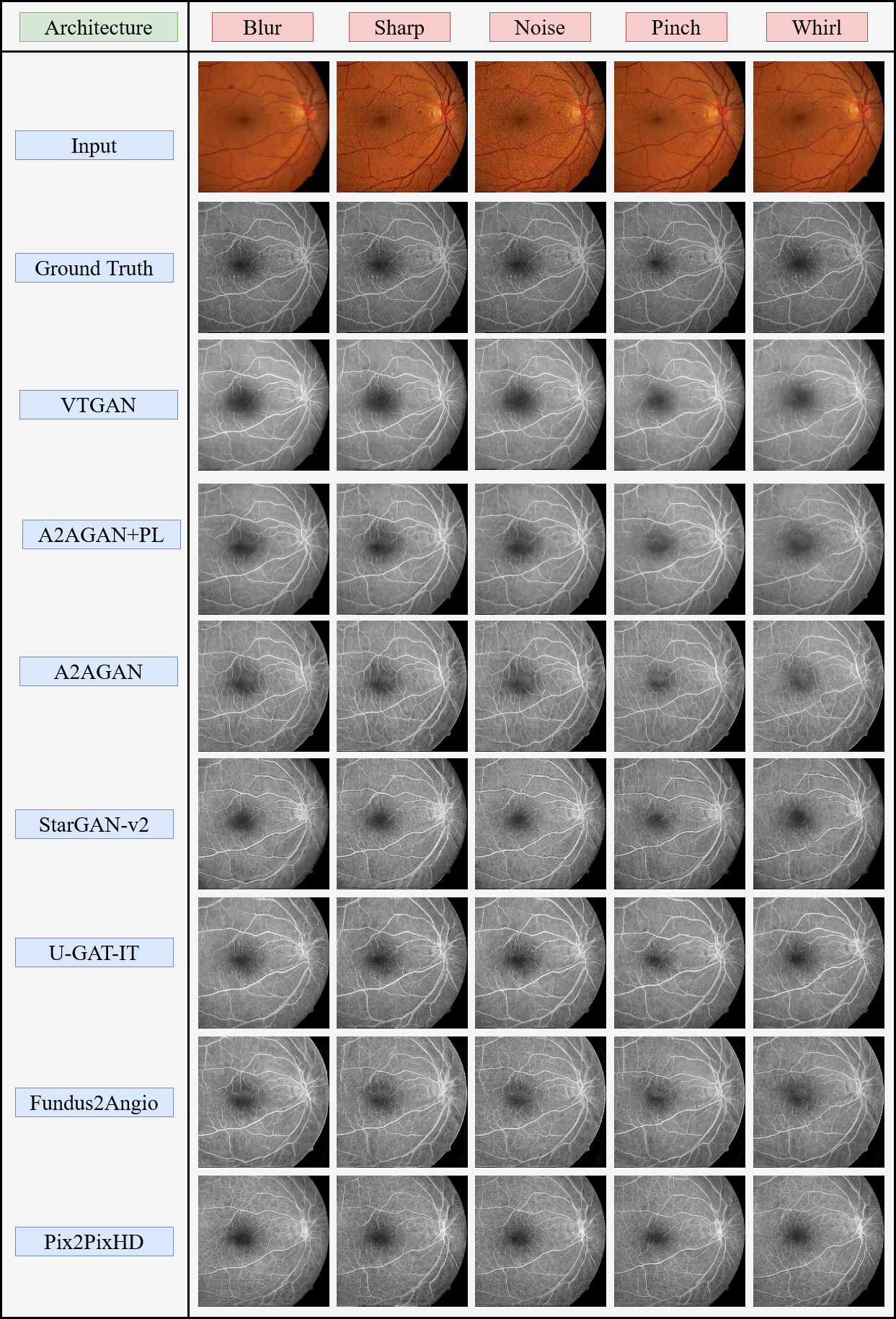}
    \caption{\textbf{1st \& 2nd rows:} Fundus and Angiogram pairs altered using spatial and radial distortions. \textbf{3rd-8th rows:} Angiograms generated from distorted fundus imitating imaging errors and distinct biological markers.}
    \label{fig4}
\end{figure} 

\subsection{Dataset}
\label{subsec:dataset}
For our experimentations, we utilize the dataset provided by \cite{hajeb2012diabetic}. The dataset comprises thirty healthy and twenty-nine unhealthy sets of fundus and fluorescein angiograms (FA). Each image pair is collected from individual patients. After close observation, we found seventeen sets that are either perfectly or closely aligned. From the original images of resolution $576\times720$, we extract 50 from each set with an overlapping crop size of $512\times512$. For the image synthesis training task, we extract 850 images. The fundus photographs are in 3 channel RGB form, and FA images are in 1 channel Gray-scale form. As the dataset is categorized into Abnormal and Normal classes, we use this annotation for our supervised classification training. Out of seventeen images, ten are Abnormal, and seven are Normal patients. Due to cropping, we end up having 500 for abnormal and 350 for normal images. We use data augmentation to increase the sample size the same as the abnormal one. To test the models, we crop four overlapping quadrants of each image pair and generate fifty-six images in total along with their associated labels. 

\subsection{Hyper-parameter tuning}
\label{subsec:hyper}
 We utilized hinge loss \cite{zhang2019self,lim2017geometric} for training our models in an adversarial setting. We chose $\lambda_{adv}=10$ (Eq.~\ref{eq4}) and $\lambda_{perc} =10$, $ \lambda_{ef} =1$,  $\lambda_{MSE} =10$, $\lambda_{cce} =10$ (Eq.~\ref{eq9}). For optimizer we used Adam \cite{kingma2014adam}, with learning rate $\alpha=0.0002$, $\beta_1=0.5$ and $\beta_2=0.999$. The batch size was, $b=2$ and we trained for 200 epochs for 50 hours with NVIDIA GPUs. Code Link : \textcolor{red}{github.com/SharifAmit/VTGAN}.

\begin{table*}[htp]
\begin{center}
\label{table1}
\caption{Test results for different architectures}
\begin{adjustbox}{width=0.8\linewidth}
\begin{threeparttable}
    \begin{tabular}{|l|c|c|c|c|c|c|} 
    \hline
    \multicolumn{7}{|c|}{\textbf{Fréchet Inception Distance (FID)}}\\
    \hline
    Architecture &  Orig. &  Noise & Blur &  Sharp & Whirl & Pinch \\
    \hline\hline
    \textbf{VTGAN}  & \textbf{17.3} & \textbf{18.7} (1.4$\uparrow$) &  24.1 (6.8$\uparrow$) &  \textbf{24.5} (7.2$\uparrow$) &  28.4 (11.1$\uparrow$) &  \textbf{22.3} (5.0$\uparrow$) \\
    A2AGAN w/ PL\tnote{1}  \cite{kamran2020attention2angiogan} & 24.6 &  21.6 (3.0$\downarrow$) &  30.0 (5.4$\uparrow$) &  25.6 (1.0$\uparrow$) &  40.0 (15.4$\uparrow$) &  24.9 (0.3$\uparrow$) \\
    A2AGAN  \cite{kamran2020attention2angiogan}& 20.7 &  20.8 (0.1$\uparrow$) &  \textbf{23.5} (2.8$\uparrow$) & 24.9 (4.2$\uparrow$) & \textbf{27.8} (7.1$\uparrow$) &  19.5 (1.2$\downarrow$) \\
    StarGAN-v2 \cite{choi2020stargan} & 27.7 & 35.1 (7.4$\uparrow$) & 32.6 (4.9$\uparrow$) & 27.4 (0.3$\downarrow$) & 32.7 (5.0$\uparrow$) & 26.7 (1.0$\downarrow$) \\
    U-GAT-IT \cite{kim2019u} & 24.5 & 26.0 (1.5$\uparrow$) & 30.4 (5.9$\uparrow$)  & 26.8 (2.3$\uparrow$) & 33.0 (9.5$\uparrow$) & 29.1 (4.6$\uparrow$) \\
    Fundus2Angio \cite{kamran2020fundus2angio} & 30.3 &  41.5 (11.2$\uparrow$) & 32.3 (2.0$\uparrow$) & 34.3 (4.0$\uparrow$) & 38.2 (7.9$\uparrow$) & 33.1 (2.8$\uparrow$) \\ 
    Pix2PixHD \cite{wang2018high} & 42.8  & 53.0 (10.2$\uparrow$)& 43.7 (1.1$\uparrow$) & 47.5 (4.7$\uparrow$) & 45.9 (3.1$\uparrow$) & 39.2 (3.6$\downarrow$) \\ 
    \hline
    \hline
    \multicolumn{7}{|c|}{\textbf{Kernel Inception Distance (KID)}}\\
    \hline
    Architecture  &  Orig. &  Noise & Blur &  Sharp & Whirl & Pinch \\
    \hline\hline
    \textbf{VTGAN} & \textbf{0.00053} & \textbf{0.04953} & \textbf{0.00205} & \textbf{0.04927} & \textbf{0.04815} & \textbf{0.04542} \\
    A2AGAN w/ PL\tnote{1}  \cite{kamran2020attention2angiogan} & 0.00087 & 0.05045 & 0.00235 & 0.05162 & 0.05390 & 0.04575 \\
    A2AGAN \cite{kamran2020attention2angiogan} & 0.00392 &	0.05390 & 0.00505 & 0.05301 & 0.05657	& 0.05341 \\
    StarGAN-v2 \cite{choi2020stargan} & 0.00118 & 0.05274 & 0.00235 & 0.05331 & 0.05539 & 0.05271 \\
    U-GAT-IT \cite{kim2019u} & 0.00131	& 0.05610 & 0.00278 & 0.05533 & 0.05815 & 0.05719\\
    Fundus2Angio \cite{kamran2020fundus2angio} & 0.00184 & 0.05328 & 0.00272 & 0.05267 & 0.05278 & 0.04985 \\ 
    Pix2PixHD \cite{wang2018high} & 0.00258 & 0.05613 & 0.00254 & 0.05788 & 0.06029 & 0.05838 \\ 
    \hline
    \end{tabular}
    \begin{tablenotes}
         \item[1] PL = Perceptual Loss
         \item[2] FID: Lower is better; KID: Lower is better
    \end{tablenotes}
\end{threeparttable}
\end{adjustbox}
\end{center}
\label{table1}
\end{table*}

\subsection{Qualitative Assessments}
\label{subsec:qual}
For evaluating our architecture's performance, we used 14 test samples and cropped four quadrants of the image with a resolution of $512\times512$. We conducted two experiments for estimating the accurate visual representation i) without transformation and ii) with spatial and radial transformations. By doing so, we measured the network's ability to adjust to structural changes due to imaging error and patient eye movements.

For our first experiment, we compare our network with other state-of-the-art image-to-image translation architectures. A side-by-side comparison of the results is visualized in Fig.~\ref{fig3}. Column, A \& C in Fig.~\ref{fig3}, display the global feature differences while column B \& D are zoomed-in local vasculature changes. By its looks, our model produces vivid and convincing results compared to other architectures. Attention2Angio (A2AGAN) and U-GAT-IT also yield impressive results. However, if observed for the closed-up versions in columns B \& D, we can witness that the optic disc contains fewer blood vessels. StarGAN-v2 and Pix2PixHD also fail to generate rich arteries, exudates, and vasculature. 

In the second set of experiments, we applied three transformations and two distortions on the fundus images: 1) Blur to represent severe cataracts, 2) Sharpening to represent dilated pupils, 3) Noise to sensor impedance during fundoscopy, 4) Pinch to visualize squinting as a  vascular change, and, 5) Whirl, for distortions caused by increased intraocular pressure (IOP). We can see in Fig.~\ref{fig4} a side-by-side comparison of different architecture predictions on these transformed images. As observed from the results, VTGAN synthesizes images very similar to the ground-truth (GT) and performs robustly to preserve the vascular structures. 

For \textbf{Blurred} fundus images, VTGAN is less impacted by the transformation compared to other state-of-the-art architecture, as seen in (row 4 to 9 of column 1) of Fig.~\ref{fig4}. The vascular structures are better retained as opposed to U-GAT-IT and Fundus2Angio. For \textbf{Sharpened} fundus, the angiogram generated by StarGAN-v2 and A2AGAN (row 4 to 6 of column 2) exhibits small artifacts around the blood vessels not present in our case. For \textbf{Noisy} images, our result is unaffected by this pixel-level alteration. However, all other state-of-the-art models (row 4 to 9 of column 3) fail to synthesize tiny and slim venular branches.

For \textbf{Pinch} and \textbf{Whirl}, our experimental result shows the versatility and reproducibility of VTGAN for retaining vascular structure and is illustrated in Fig.~\ref{fig4} (row 3 of column 4 and 5). Compared to ours, only A2AGAN and U-GAT-IT reserves the flattening condition and manifestation of vascular changes but loses the overall smoothness in the process (row 5 to 7 of columns 4 and 5). In Fig.~\ref{fig4} VTGAN encodes the vascular feature information and is much less affected by both kinds of warping. The other architectures failed to generate microvessels due to IOP or vitreous variations, as can be seen in Fig.~\ref{fig4}. Consequently, For all kinds of transformation and contortion, VTGAN surpasses existing state-of-the-art image-to-image translation models.

\subsection{Quantitative Assessments}
\label{subsec:quant}

For quantitative evaluation, we performed two experiments.  In the first experiment, we use the Fréchet inception distance (FID) \cite{heusel2017gans}, and Kernel Inception distance (KID) \cite{binkowski2018demystifying} which has been previously employed for measuring similar image-to-image translation GANs \cite{choi2020stargan,kim2019u}. We computed the FID and KID scores for different architectures on the generated FA image and original angiogram, including the five spatial and radial transformations. The results are given in Table.~1. A lower FID and KID score means the synthesized images are more close to the real angiogram. 

From Table.~1,  out of our three networks, the best KID is achieved for VTGAN.  And it reaches the lowest scores out of all other architecture, for both with and without distortions. For FID, our model achieves the lowest score for three out of five types of distortions. Attention2Angio scores lower FID for Blur and Whirl transformation. 

In the second experiment, we assess the generated angiogram's quality by asking two expert ophthalmologists to identify them. We use a balanced set of 50 images, 25 real and 25 fake. We then shuffle the data before the expert evaluation. For this experiment, experts did not know the exact number of fake and real images. By not disclosing this, we tried to estimate the following criterion: 1) Correct fake and real angiograms found by the experts, where lower equates to better, 2) Incorrect fake and real angiograms missed by the experts, where higher equates to better, and 3) The average precision of the experts for identifying fake angiograms, where lower equates to better. Table~\ref{table2} illustrates the detailed result.

As it can be seen from Table~\ref{table2}, experts assigned 94\% of the fake angiograms as real, for images synthesized by our models. The result also shows that experts had trouble identifying fake images, while they easily identified real angiograms with 80\% certainty. On average, the experts misclassified 57\% of all images produced by VTGAN. The average precision diagnosis of the experts are 45.9\%. Consequently, our model successfully fools the experts to identify fake angios as real.

\begin{table}[t]
\caption{Results of Qualitative with Undisclosed Portion of Fake/Real Experiment}
\label{table2}
\centering
\begin{adjustbox}{width=\linewidth}
\begin{threeparttable}
\begin{tabular}{|l|l|c|c|c|c|c|} 
\hline
&&\multicolumn{2}{c|}{\small Results} & \multicolumn{3}{c|}{ \small Average}  \\\hline
Architecture && \small Correct & \small Incorrect & \small Missed\tnote{1} & \small Found\tnote{1} & \small Precision\tnote{2}\\\hline\hline
\small \multirow{2}{*}{VTGAN} &\small Fake & \small 6\% & \small 94\% & \small \multirow{2}{*}{57\%} & \small \multirow{2}{*}{43\%} & \multirow{2}{*}{\small \textbf{45.9\%}} \\
&\small Real & \small 80\% & \small 20\% & & & \\
\hline
\end{tabular}
    \begin{tablenotes}
         \item[1] Missed higher is better; Found lower is better
         \item[2] Precision Lower is better
    \end{tablenotes}
\end{threeparttable}
\end{adjustbox}
\end{table}

\begin{table}[bp]
    \centering
    \caption{Test Accuracy on in-distribution Abnormal/Normal Angiograms}
    \begin{tabular}{|c|c|c|}
        \hline
     \textbf{Accuracy}  & \textbf{Sensitivity} & \textbf{Specificity}\\
    \hline\hline
                85.7 & 83.3 & 90.0\\   
        \hline
    \end{tabular}
    \label{table3}
\end{table}

\subsection{Disease Classification}
\label{subsec:classification}
For our subsequent experimentation, we test our visual transformer's accuracy on in-distribution and out-of-distribution classification tasks. We use data provided by Hajeb \etal \cite{hajeb2012diabetic}. We crop four quadrants from the 14 Fundus and Angiogram pairs because of the shortage of data. Out of 56 images, 20 are for Abnormal, and 36 are for Normal classes. We name this test set as in-distribution and measure the performance using three standard metrics: Accuracy, Sensitivity, and Specificity. The result is provided in Table.~\ref{table3}, and our vision transformer-based model scores 85.7\%, 83.3\%, 90\% for accuracy, sensitivity, and specificity successively.

We use spatial and radial transformation on test images for out-of-distribution evaluation. The model's performance is illustrated in Table.~\ref{table4}. As can be seen from the table, for Blur, Noise, Pinch, and Whirl transformations, the Accuracy and Sensitivity decreased to 78.6\%, 72.2\%, consecutively. Compared to the in-distribution data, the decrease is 7.1\%, 11.1\% for Accuracy and Sensitivity. For Sharp transformation, the accuracy and sensitivity are worse than the other distortions. They are  76.7\%, 69.4\% successively. An interesting revelation is the specificity has no effect due to these distortions. The specificity is firm 90.0\% for all five distortions. A higher specificity signifies our model accurately predicts Abnormal classes better compared to Normal ones. This is significant, as we want to identify patients with degenerative conditions compared to predicting false positives for healthy patients.

\begin{table}[t]
    \centering
    \caption{Test Accuracy on out-of-distribution Abnormal/Normal Angiograms}
    \begin{tabular}{|c|c|c|c|}
        \hline
    \textbf{Distortion} & \textbf{Accuracy}  & \textbf{Sensitivity} & \textbf{Specificity}\\
    \hline\hline
                Blur &   78.6 (7.1 $\downarrow$) & 72.2 (11.1 $\downarrow$) & 90.0 (-) \\
                Sharp &  76.7 (8.0 $\downarrow$) & 69.4 (13.9 $\downarrow$) & 90.0 (-) \\
                 Noise & 78.6 (7.1 $\downarrow$) & 72.2 (11.1 $\downarrow$) & 90.0 (-)\\
                 Pinch & 78.6 (7.1 $\downarrow$) & 72.2 (11.1 $\downarrow$)  & 90.0 (-)\\
                 Whirl & 78.6 (7.1 $\downarrow$) & 72.2 (11.1 $\downarrow$)  & 90.0 (-)\\         
        \hline
    \end{tabular}
    
    \label{table4}
\end{table}

\section{Conclusion}
In this paper, we proposed a new fundus-to-angiogram translation architecture called VTGAN. The architecture generates realistic angiograms from fundus images without any expert intervention. Additionally, we provided results for its robustness and adaptability conditioned upon radial and spatial transformations, which imitate biological markers seen in real fundus images. We believe the proposed network can be incorporated in the wild to generate precise FA images of patients developing disease overtime. It can be a complimentary disease progression monitoring system for predicting the development of diseases in vivo. We hope to extend this work to other areas of ophthalmological image modalities. 

\section*{Acknowledgement}
This material is based upon work supported by the National Aeronautics and Space Administration under Grant No. 80NSSC20K1831 issued through the Human Research Program (Human Exploration and Operations Mission Directorate). 

{\small
\bibliographystyle{ieee_fullname}
\bibliography{egbib}
}

\end{document}